\documentclass[aip,10pt,reprint]{revtex4-1}
\usepackage{amsmath}    
\usepackage{graphicx}   
\usepackage{verbatim}   
\usepackage{color}      
\usepackage{subfigure}  
\usepackage{hyperref}   
\usepackage{wasysym}
\usepackage[utf8]{inputenc}

\begin{document}
\title{Performance of the new 2D ACAR spectrometer in Munich}
\author{Hubert Ceeh}
\email{hubert.ceeh@frm2.tum.de}
\affiliation{Technische Universität München, Lehrstuhl E21, James-Franck Straße, 85747 Garching, Germany }
\author{Josef-Andreas Weber}
\affiliation{Technische Universität München, Lehrstuhl E21, James-Franck Straße, 85747 Garching, Germany }
\author{Michael Leitner}
\affiliation{Technische Universität München, Lehrstuhl E13, James-Franck Straße, 85747 Garching, Germany }
\author{Peter B\"oni}
\affiliation{Technische Universität München, Lehrstuhl E21, James-Franck Straße, 85747 Garching, Germany }
\author{Christoph Hugenschmidt}
\affiliation{Technische Universität München, Lehrstuhl E21, James-Franck Straße, 85747 Garching, Germany }
\affiliation{FRM\,II, Technische Universit\"at M\"unchen, Lichtenbergstra\ss e 1, 85747 Garching, Germany}
\date{\today}

\begin{abstract}
Angular Correlation of Annihilation Radiation (ACAR) is a well established technique for the investigation of the electric structure.  
A major limitation of ACAR studies is the available positron flux at a small spot on the sample. 
Fore this reason, the focus of this work is put on the discussion of a newly developed  source-sample stage which uses an optimized static magnetic field configuration to guide the positrons onto the sample. 
The achieved spot size is $d_{\mathrm{FWHM}}=5.4\,$mm, with a high efficiency over the whole energy spectrum of the $^{22}$Na positron source.
The implications of the performance of the source-sample stage are discussed with regard to 2D-ACAR measurements of single crystalline $\alpha$-quartz, which serves as a model system for the determination of the total resolution.
\end{abstract}
\maketitle
\section{Introduction}
All properties of materials are defined by their electronic structure. For a metal, the most important characteristic is the boundary between occupied and unoccupied states in reciprocal space, which is called the Fermi surface. In fact, a wide range of phenomena can be explained in terms of its shape and topology and have been investigated by 2D-ACAR, e.g. ferromagnetism\cite{Nickel91}, superconductivity\cite{Supra92,YBaCuO88}, charge- and spin-density waves\cite{Spin04}, ferromagnetic shape-memory effect\cite{Dug12}, or large effects of small compositional changes\cite{Comp89,Eijt2006}.
\\
The traditional experimental method for studying the Fermi surface is the de Haas-van Alphen effect, which records the response of the sample to high external magnetic fields, however, it requires low measurement temperatures and highly ordered systems.   
The method of Angle-Resolved Photo-Emission Spectroscopy (ARPES), on the other hand, is only sensitive to surfaces, i.e. topmost few atomic layers\cite{ARPES03}. Both methods therefore cannot serve as a general-purpose tool for studying effects such as bulk phase transitions. \\
Angular Correlation of Annihilation Radiation (ACAR) from positron electron annihilation in matter is free of limitations such as low temperatures and high magnetic fields, therefore temperature dependent studies of the electronic structure are feasible. 
ACAR works a follows: in ACAR experiments the deviation from the anti-parallel propagation directions of the two annihilation quanta is measured in coincidence. 
Using spatially resolved detectors, two dimensional projections of the electron density can be measured, which are used to reconstruct the three dimensional electron momentum density of the sample.
For this method, the limitations lie 
 in the achievable count rate and resolution. 
Therefore, the goal is to guide by means of a magnetic field as many of the emitted positrons (from a radioactive source) with a wide spectrum of energies onto a sample within a small spot as small as possible since this defines the achievable momentum resolution. \\
Here we present an overview of the recently completed 2D-ACAR spectrometer and the design of a source-sample stage optimized for efficient positron transport as well as the measured performance. 

\section{Layout of the new 2D-ACAR spectrometer}
\subsection{Overview} 
The 2D-ACAR spectrometer consists of two main components, the source-sample chamber including the sample environment, and the detector assembly consisting of two Anger-type $\gamma$-cameras \cite{Anger}, which were obtained from the positron group of Bristol university. A total view of the spectrometer is given in figure \ref{fig:over} (A). 
\subsection{Detector system}
The Anger cameras consist of a large NaI:Tl scintillation crystal with a thickness of $10.5\,$mm coupled to 61 photomultiplier tubes. The active area is collimated to a diameter of $41.5\,$cm by a lead ring, in order to exclude the part of the detector, where the  position response becomes non-linear, due to the discontinuity in the light collection efficiency at the borders of the crystal.
The position information is obtained from  the center of mass of the light signal produced in the scintillator.
The individual photomultiplier signals are summed up in an analogue weighting resistor network in horizontal and vertical direction and are then divided by the integral signal of all photomultiplier tubes. 
The resulting x- and y-signals are shaped and amplified before they are fed into the data acquisition system. Parallel to the determination of the center of mass a logic analyzer circuit checks if the integral signal, which corresponds to the energy of the event, lies within a window of about $\pm \,35keV$ around the $511\,$keV photo peak. 
If this is the case an logic signal is produced and is fed to an analogue coincidence unit. If both cameras produce an appropriate analyzer signal, the data acquisition is triggered by the coincidence unit.\\

The two Anger cameras are positioned symmetrically at a distance of $8.25\,$m to the source-sample chamber. 
As the typical angular correlation lies in the order of milliradians, a long baseline is needed to resolve such small angles. 
The angular resolution is limited due to the finite spatial resolution of $\approx 3.5\,$mm of the Anger cameras. 
By increasing the baseline the angular resolution could be improved at the expense of the count rate which  decreases quadratically with the length of the baseline. 
To avoid errors by a transverse pitch and yaw of the Anger cameras their respective axis' are aligned collinear with an accuracy of $\pm1\,$mm  using a laser positioning system in combination with a theodolite. 
In the same way it is made sure that the  sample plane lies on the detector axis.
In order to damp temperature fluctuations in the experimental hall the Anger cameras are housed inside acrylic glass boxes. 
 
 \subsection{Sample environment}
 The new 2D-ACAR spectrometer features two interchangeable sample holders, for both heating and cooling the sample.  
 In principle temperatures between $\approx 5\,$K and $500\,$K can be achieved depending on the sample and hence enable the study of temperature driven effects on the electronic structure. \\
  The coolable sample holder consists of a closed-cycle cryostat with an extended $22\,$cm long cold finger. The cooling power is controlled with a PID-controller via two $50\,\Omega$ cartridge heaters and a Si diode that are directly coupled to the 2nd stage of the cryostat. Since the temperature at the sample position is slightly different, the sample temperature is separately monitored with an additional Si diode which is mounted close to the sample.\\
 The heatable sample holder is comprised of a heating filament, which is wound on a copper rod to increase the inertia of the system. The temperature is measured with a type-K thermo-couple that is tightly fastened to the copper rod and controlled with a standard PI-controller. To minimize the necessary heating power the heater block is decoupled from the vacuum chamber by stainless steel fittings with low thermal conductivity. 
\subsection{Design of the source-sample chamber}
\begin{widetext}
    		\begin{figure}[ht]
		\begin{center}
			\includegraphics[width=0.90\textwidth]{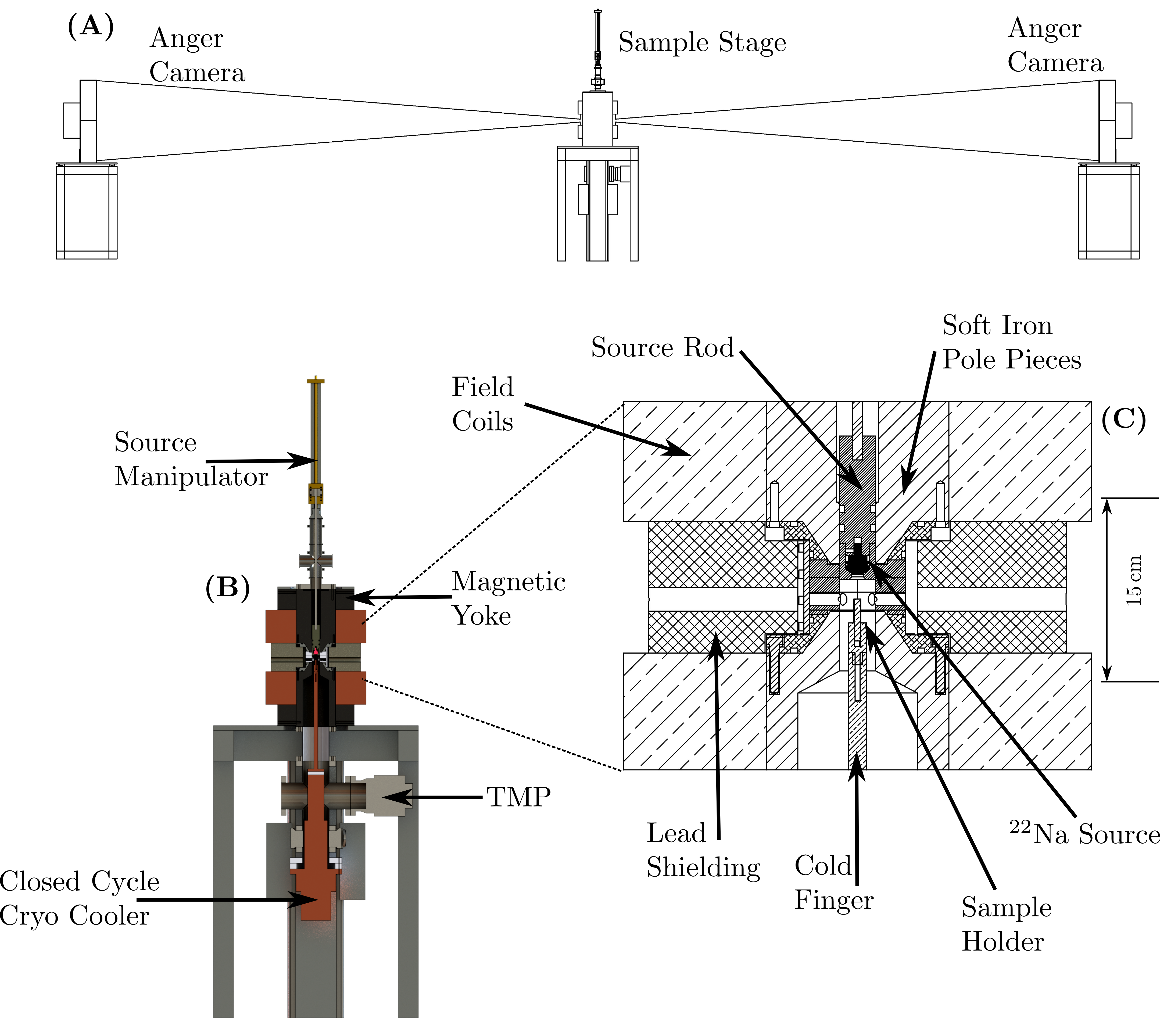}
		\end{center}
		\caption{\label{fig:over} Schematic overview of the ACAR spectrometer. (A)Total view of the spectrometer, including detectors. The baseline of the anger cameras is is $16.5\,$m in order to achieve the desired angular resolution. (B) Cut view of the central assembly. The cold finger of the cryostat is fed into the sample chamber from below. (C) Detail cut-view of the source-sample chamber with the pole piece assembly. Sample and source are positioned  symmetrical with respect to the pole pieces. }
		\end{figure}
\end{widetext}

The challenges for ACAR lie on the one hand in the maximization of the count rate, and on the other hand in the optimization of the achievable resolution, which is given by the spatial resolution of the detectors, the sample temperature and the positron spot size on the sample. 
The detector resolution, which is given by the maximum CSDA-range of the photo electron in NaI of $0.62\,$mm\cite{Estar} and the position dependent weighting of the scintillation light by the photomultiplier read-out, can not be improved easily. 
Optional sample cooling would reduce the smearing-effect of the positron momentum and hence leads to a higher resolution. The resolution smearing due to  the positron motion ranges between $\sigma_{\mathrm{therm}}=0.12\,$mrad at low temperatures ($<10\,$K) and $\sigma_{\mathrm{therm}}=0.67\,$mrad at room temperature\cite{Kub83}.
In order to improve the resolution our main concern is to minimize the positron spot on the sample. 
In addition, the sample should be easily accessible, and the spectrometer should enable  the study of temperature dependent effects on the electronic structure, i.e. phase transitions.\\
We took all these considerations into account when designing the setup presented in figure \ref{fig:over} (B) and (C). The positrons are emitted from a $^{22}$Na source deposited on a Ta reflector inside a standard source capsule.
 The Ta reflector is used to increase the emission of positrons into the lower half space. 
The source capsule is held inside an elkonite rod, which is connected to a manipulator. So the source can be moved inside a heavily shielded storage position, for example when the sample is changed. The position of the source capsule and the sample are symmetric with respect to the pole pieces of the electromagnet. For all considerations presented below a distance between source and sample of $20\,$mm was chosen, as this value turned out to be optimal with regard to the background produced by the source itself and the positron transport from the source to the sample. 
A larger distance between source and sample would decrease further the background in the ACAR spectrum, but puts higher requirements to the magnetic guiding field.\\
To produce the guiding field two custom build soft-iron pole pieces are used. 
The layout of the pole pieces was developed using the FEM physics simulation toolkit COMSOL\citep{Comsol}. 
Diameter and height of the truncated cone shaped pole pieces were optimized in a way that a region with a homogeneous ($<1\%$ in the direction transverse to the axis between source and sample) flux density is created around sample and source (see figure \ref{fig:femsimu}). 
The pole pieces have a central bore which allows for the retraction of the source from the sample chamber and for feeding the cold finger of a 4K closed-cycle cryocooler into the sample chamber. The bore in the base of the lower pole piece is widened so that it can accommodate the heat shield of the cryocooler.  
The central sample chamber connects the two pole pieces at a fixed distance and also makes them part of the vacuum system. The two pole pieces are therefore  an integral part of the sample chamber, which is pumped by a turbo molecular pump from below. This layout frees up space in the central part between the field coils and hence allows the lead shielding to be more compact. The pole piece assembly is placed inside a commercially available  electro magnet with a soft iron yoke. A $5\,$kW high current power supply ($I_{\mathrm{max}}=75\,$A) is used to generate the magnetic field of up to $1.1\,$T.\\
The cryocooler is attached to the vacuum system by a differentially pumped rotary platform. This way the orientation of the sample can be changed by $\pm 90^{\circ}$ with respect to the detector axis without breaking the vacuum or cooling.\\
The annihilation radiation  produced in the sample leaves the sample chamber trough $1\,$mm thin aluminium  windows. The transmission of these windows is $97.7\,\%$ for $511\,$keV quanta. In order to screen the background contribution from the source, the sample chamber is enclosed by lead shielding with a $16\,$mm bore along the line of sight from the sample to the detectors. 
Therefore, apart from the line of sight of the detectors the radiation exposure in the lab is minimized. 
%

%
\subsection{Magnetic field configuration}
\begin{figure}[htbp]
  
  \subfigure[\label{fig:femsimu}]{
    \includegraphics[width=0.45\textwidth]{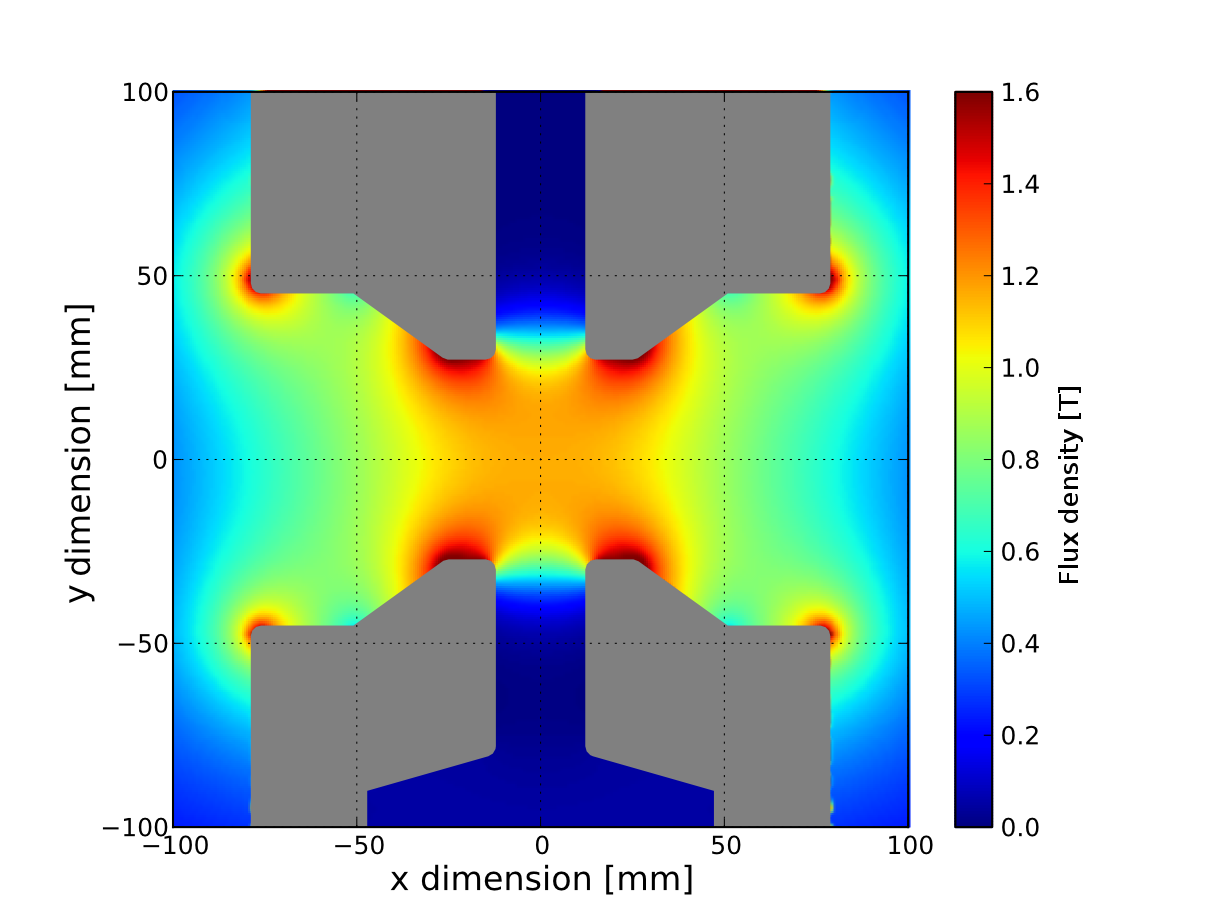}
  }
  \subfigure[\label{fig:measFeld}]{
    
    \includegraphics[width=0.45\textwidth]{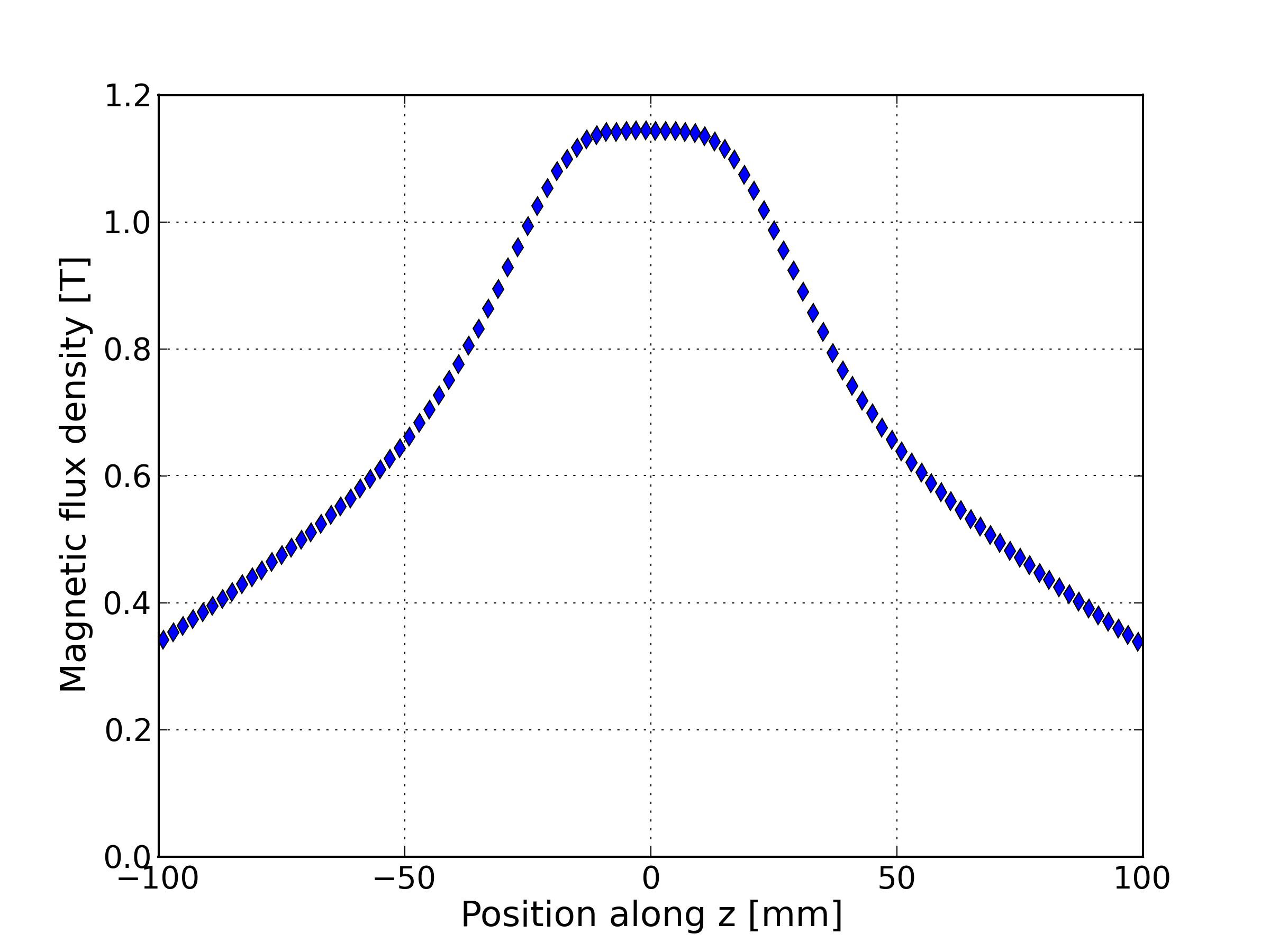} 
  }
  \caption{ \label{fig:Feld}Magnetic field configuration. \ref{fig:femsimu} FEM Simulation of the magnetic flux density inside the sample chamber. Between the two pole pieces a region with a homogeneous field is created. Sample and source are placed inside this region for optimal positron transport. \ref{fig:measFeld} Measured magnetic field in sample plane. The sample position is located on the plateau the center. The flux density decreases rapidly outside the pole piece configuration.}
 
\end{figure}

\begin{figure}[h]
\includegraphics[width=0.40\textwidth]{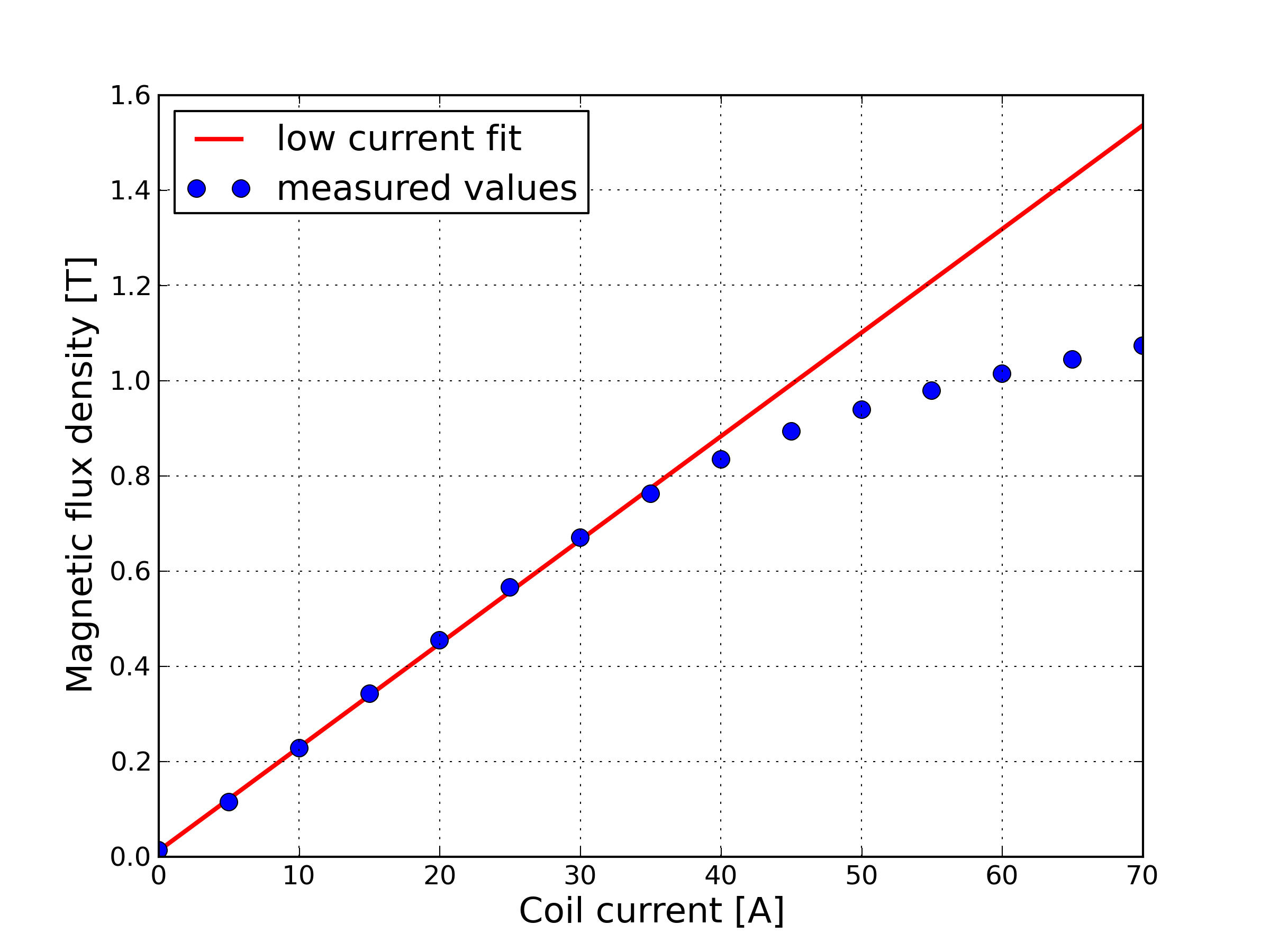}
\caption{\label{fig:strom} Dependence of the magnetic flux density at the sample position on the electric current through the coils. At a current of $I_{\mathrm{coil}}\approx 45\,$A the soft iron is beginning to saturate.}
\end{figure}

%
%
The field configuration was measured with a hall probe mounted on a 3-axis stepper motor driven manipulator. 
In this way, the field throughout the entire sample chamber was mapped out and compared to the FEM simulation shown in figure \ref{fig:femsimu}. Excellent agreement is found concerning the shape of the field distribution. Exemplary a central line scan trough the chamber in the sample plane is shown in figure \ref{fig:measFeld}. The plateau is centred around the sample position and has a  diameter (at 0.98$\cdot\,B_{max}$) of $20\,$mm, which is well above the typical sample diameter of $6-8\,$mm. The strength of the magnetic stray field drops rapidly outside the magnetic yoke to values $<1\,$mT at a distance of more than $40\,$cm.\\
The maximal achievable flux density is limited by the current through the coils, since the distance of the pole pieces is fixed. Using water cooled copper coils a flux density of up to $1.2\,$T can be achieved. However, during routine operation a value of $1.0\,$T is chosen, as this greatly reduces the heat dissipation in the coils. Also, the soft iron is already beginning to saturate at such high fields (see figure \ref{fig:strom}).
\section{Performance}
The overall performance of a 2D-ACAR spectrometer can be judged by two figures of merit: count rate and resolution. 
In order to minimize statistical uncertainties, typically $10^8$ events are collected in each spectrum.
Therefore, the measurement time is determined by the achieved count rate, which is correlated to the activity of the source and the positron transport efficiency from the source to the sample. 
The resolution of the spectrometer is, among other contributions, limited by the spot size on the sample.

\subsection{Positron transport efficiency}
\begin{figure}[htbp]
  \subfigure[\label{fig:specfeld}]{

    \includegraphics[width=0.45\textwidth]{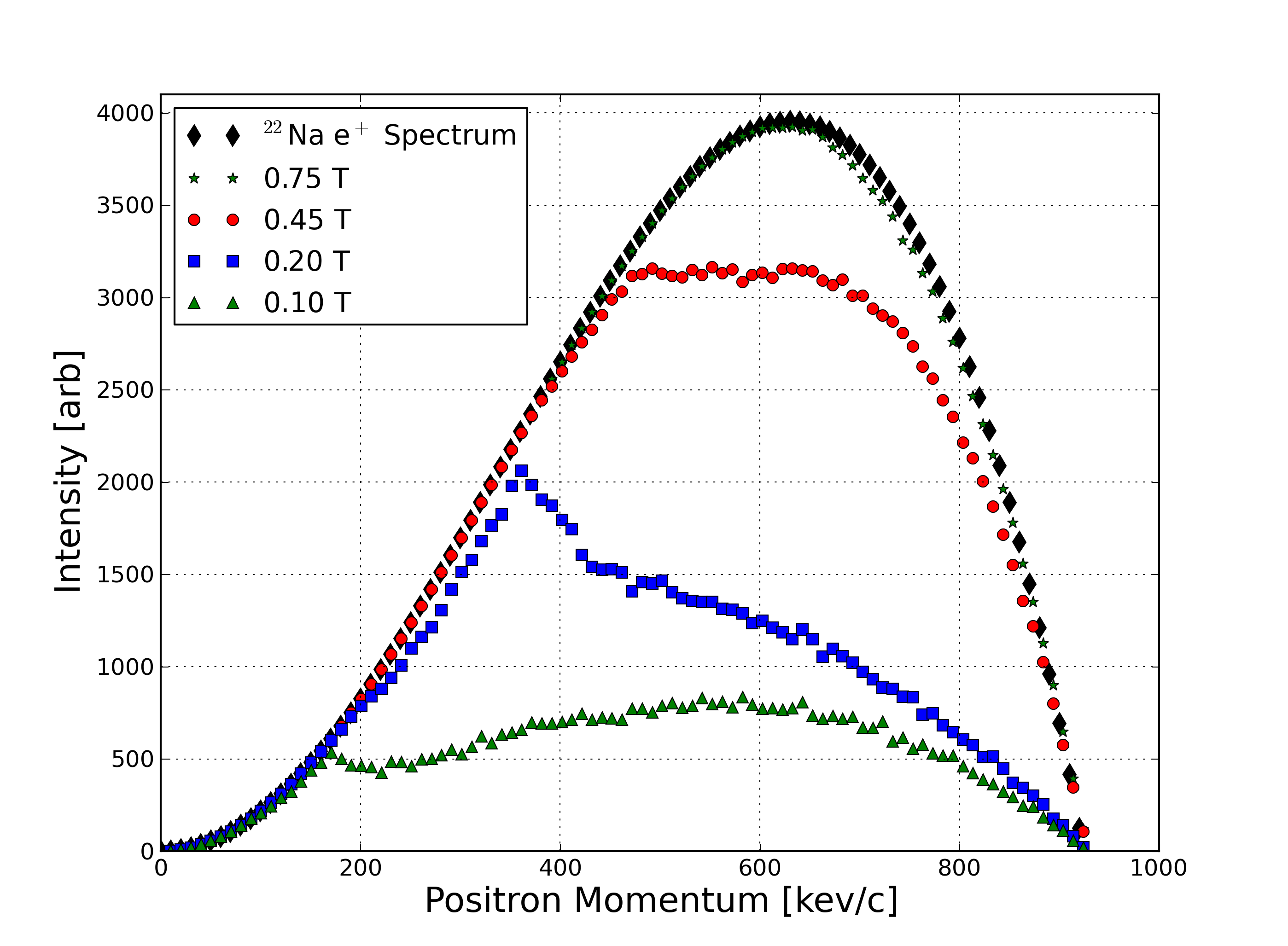} 
  }
  \subfigure[\label{fig:transSimu}]{
    \includegraphics[width=0.45\textwidth]{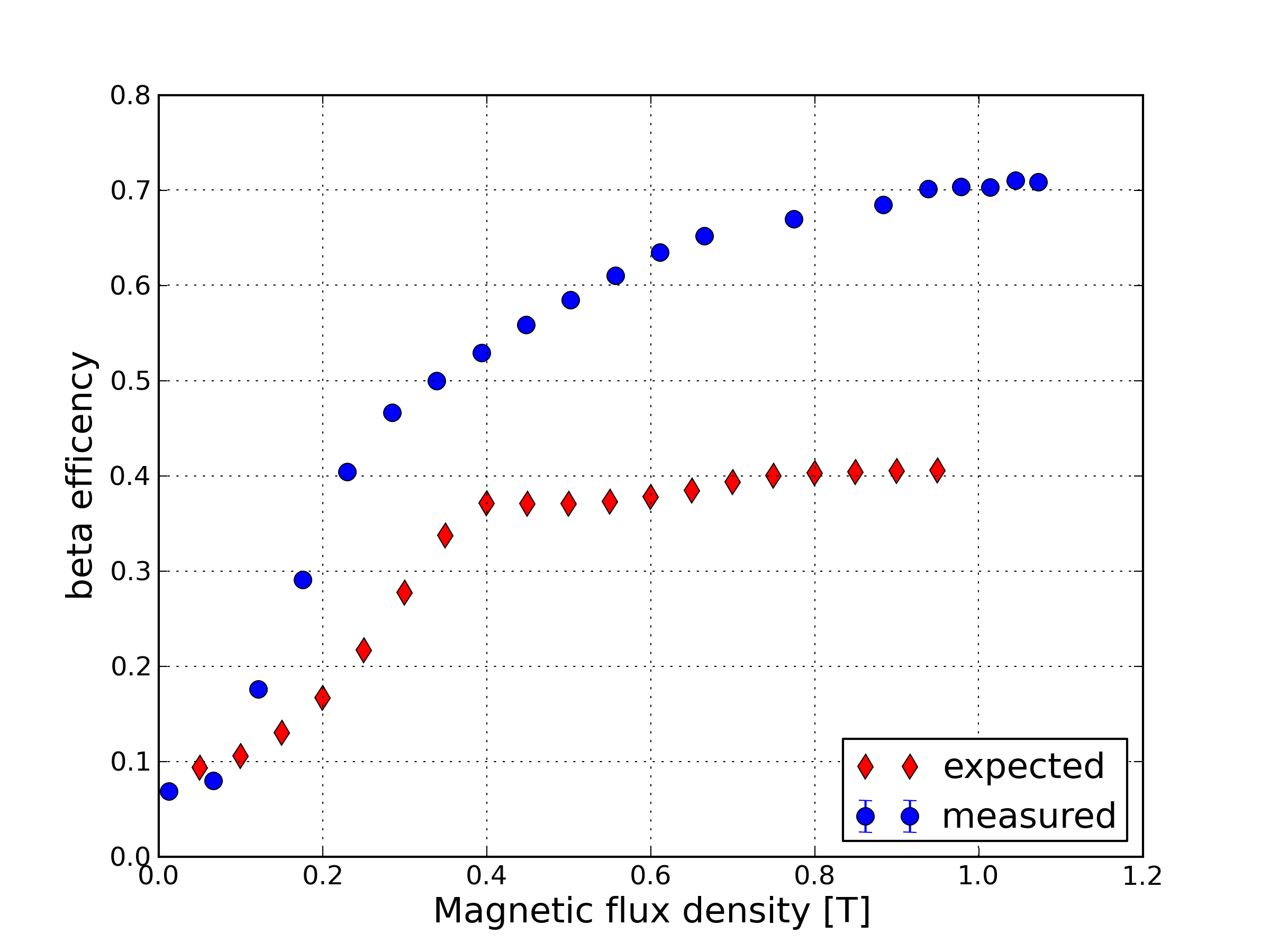}
  }
  \caption{Comparison between Monte-Carlo-Simulation and measurement for the field dependent transport efficiency for a Al sample with a diameter of $ 10\,$mm. \ref{fig:specfeld} Calculated positron momentum distribution on the sample, obtained with a Monte-Carlo simulation. For high flux densities almost the full spectrum is mapped onto the sample. \ref{fig:transSimu} Measured beta efficiency in comparison with the expectation derived from the simulation. The measured values exceed the expectation indicating that positrons which are emitted into the upper semi space are effectively reflected downwards onto the sample. }
\label{fig:trans}
\end{figure}

The coincident count rate $C_r$ in the detectors is given by the source activity $A$, the branching ratio $B_r$ for $\beta^+$-decay of the source nuclide, the detector efficiencies  $\epsilon_1$ and $\epsilon_2$, the solid angle $\Omega$, the transmission $\eta$ for $511\,$keV $\gamma$-radiation and the positron transport efficiency $\beta$: 
\[
C_r=2\cdot A \cdot \beta \cdot B_r \cdot  \eta^2 \cdot \epsilon_1 \cdot \epsilon_2 \cdot \Omega.
\]
The positron transport efficiency accounts for the probability of a positron, which is produced in the source, to hit the sample. For this estimation the deviation from collinearity of the annihilation quanta is not considered, as only the edge region of the detectors is affected. 
The individual efficiencies of the detectors can be easily determined by the ratio of the individual single count rates and the coincident count rate. The  efficiencies values are found to be $\epsilon_1=(7.4\pm0.1)\% $and $\epsilon_2 =(6.6\pm 0.1)\%$. The source activity  at the time of this study was $A=(1.56 \pm 0.05)\,$GBq The branching ratio for $\beta^+$-decay of $^{22}$Na is $90.3\%$ \cite{Na22}. The factor of 2 has to be considered, as the two annihilation $\gamma$ rays cannot be distinguished. Also the absorption $1-\eta=18.6\,\%$ of $511\,$keV quanta in the chamber windows ($5\,$mm in this case) and in the air between the chamber and the detectors has to be taken into account.  Finally the solid angle is given by distance between the detectors and the sample and the active area of the detectors $\Omega=1.58\cdot 10^{-4}$. \\
For the target a low-$Z$-material was chosen to minimize the effect of positron reflection, in this case a polycrystalline aluminium disc with a diameter of $d=10\,$mm. With a magnetic flux density of about $1\,$T a coincident coincident rate of $C_r=1.0\cdot 10^3\,\mathrm{s}^{-1}$ was achieved. 
Therefore, the positron transport efficiency can be calculated to $\beta=69\%\pm 4\%$. \\
This value, meaning  that essentially two of three produced positrons hit the sample, appears surprisingly high, as the positron emission from $^{22}$Na is isotropic. 
However, only positrons, which are emitted into the lower half space can be guided onto the sample, and self absorption inside the source would lead to a reduced positron yield at the sample. \\
In order to compare measured positron intensities with calculated ones Monte-Carlo simulation were performed for various magnetic guide fields.
The momentum distributions of positrons, which hit the sample are shown in figure \ref{fig:specfeld} together with the full spectrum from the source. 
In figure \ref{fig:transSimu} the measured count rate on a  aluminium disc with a diameter of $d=10\,$mm  is compared to the expected count rate if only emission of positrons into the lower half space is considered. As can bee seen, the measured count rate exceeds the expectation for magnetic flux densities higher than $0.1\,$T. This can be understood in terms of the Ta backening of the source, as this increases the positron emission into the lower half space.

\subsection{Determination of the positron spot size}
\begin{figure}[t]
\includegraphics[width=0.45\textwidth]{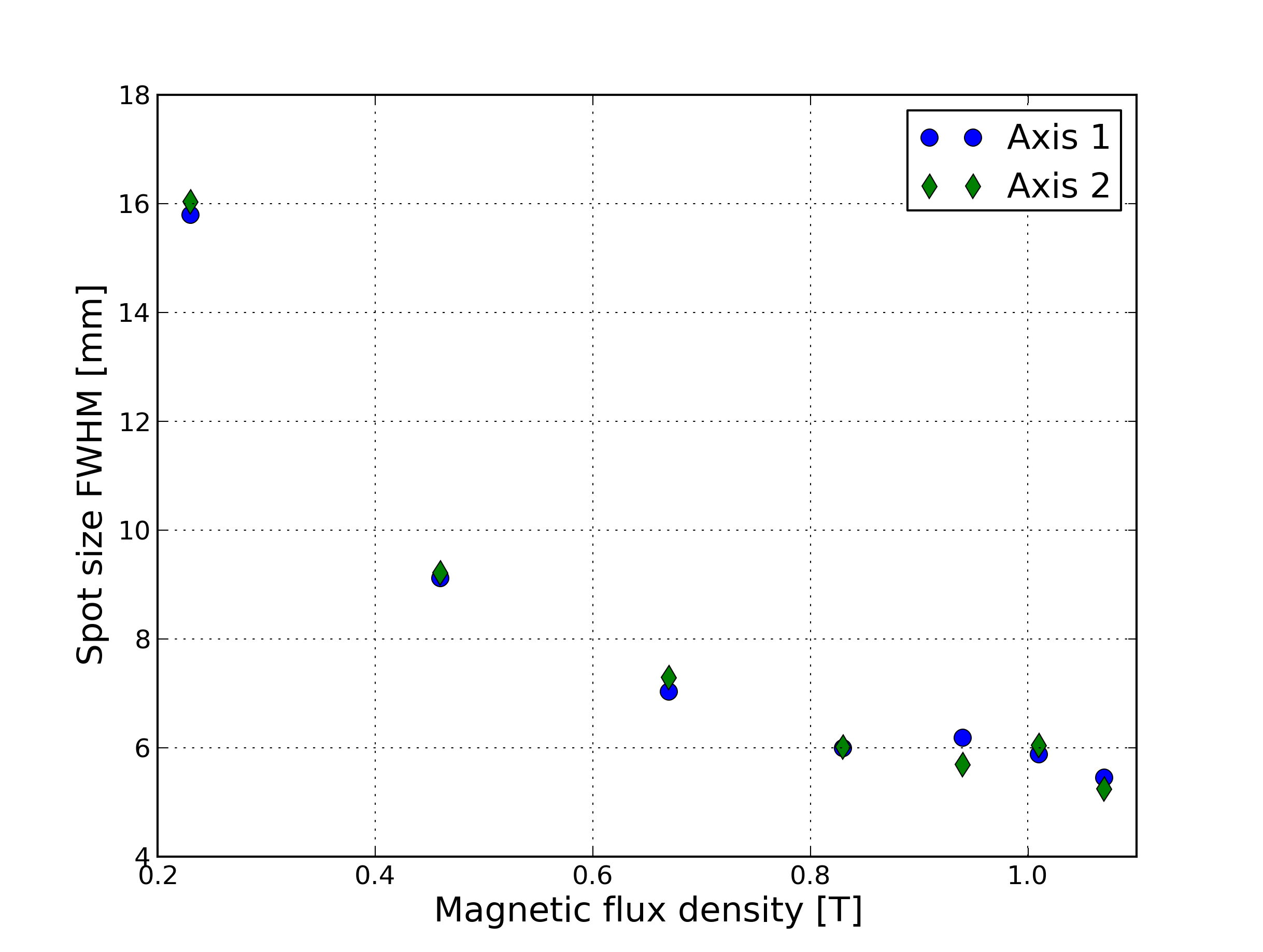}
\caption{\label{fig:spot} Dependence of the spot size FWHM on the magnetic flux density at the sample position for two arbitrary orthogonal axis.}
\end{figure}

The size of the positron spot on the sample has been measured by moving a guillotine-shaped aluminium target sheet, which was mounted on a stepper-motor driven linear translation stage, through the opened sample chamber. 
The annihilation radiation was measured with a bismuth-germanate scintillation detector that was collimated to the target. 
By recording the count rate as a function of the target position the spot profile can be estimated by taking the derivative of the measured integral distribution. 
In this way two orthogonal cuts through the spot profile could be measured simply by flipping the guillotine-shaped target, without having to change the direction of the translation stage \cite{Hug02}. \\ 
 The measurement was performed for different magnetic fields. The cuts are found to have a profile that can be described by a single Gaussian. 
 An absolute error of $\pm 0.35\,$mm is assumed for limited positioning accuracy of the translation stage. In combination with the statistical error from the fitting procedure a total error of $0.5\,$mm appears reasonable. 
 Within the estimated error the values for the FWHM in the two directions are compatible (see figure \ref{fig:spot}). 
 The resulting spot diameter (FWHM) at typical operation conditions ($1\,$T) is found to be $(5.4 \pm 0.5)\,$mm. 
 Hence, the contribution to the angular resolution in x-direction is given by $(0.65\pm 0.06)\,$mrad for a sample-detector distance of $8.25\,$m.
\subsection{First 2D-ACAR measurements}
\begin{figure}
 \includegraphics[width=0.45\textwidth]{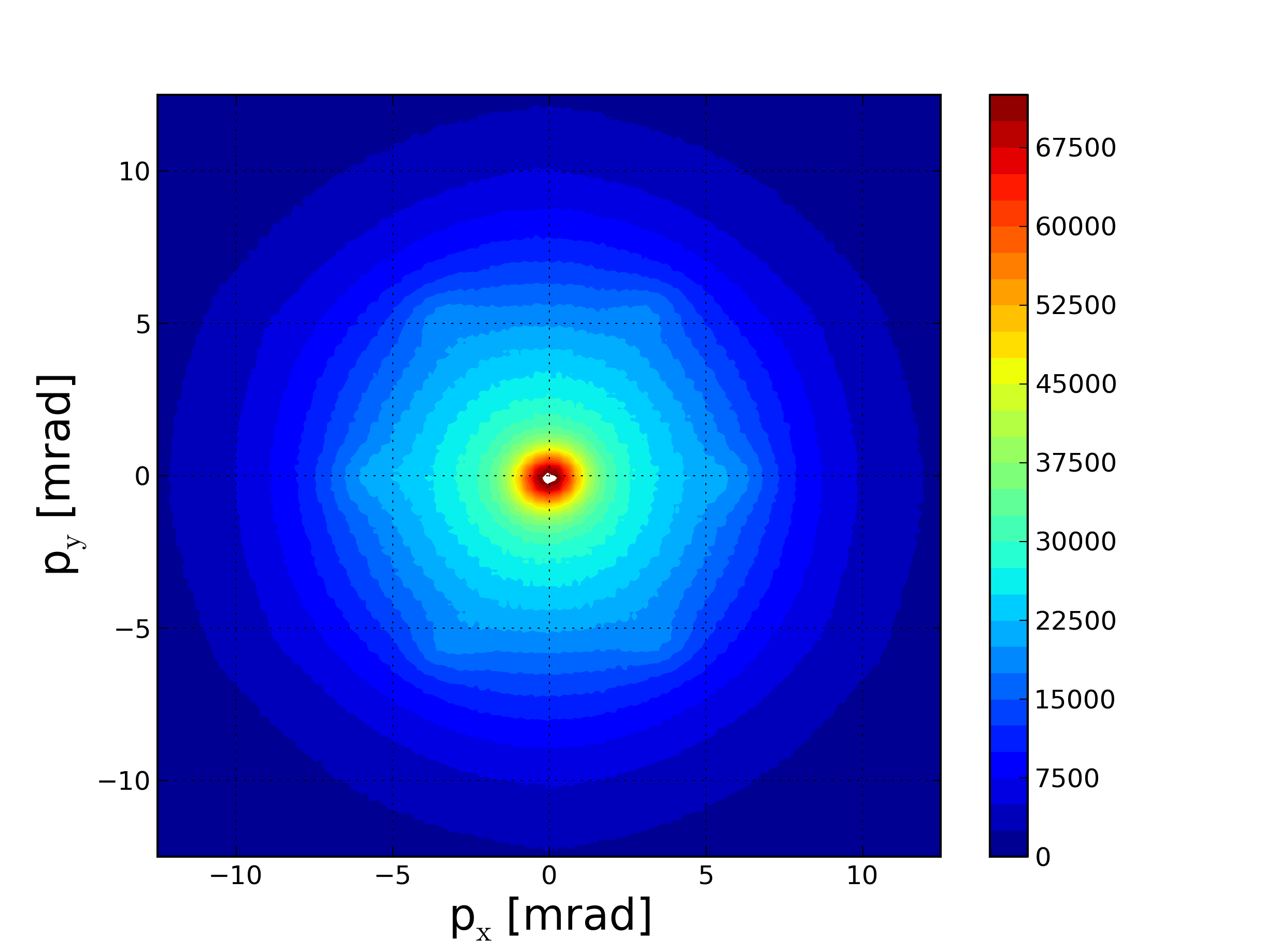}
\caption{\label{fig:quarz} Contour plot of the ACAR spectrum of single crystalline $\alpha$-Quartz}
\end{figure}

\begin{figure}[htbp]
  \subfigure[horizontal cut]{
    \label{fig:hor}
    \includegraphics[width=0.45\textwidth]{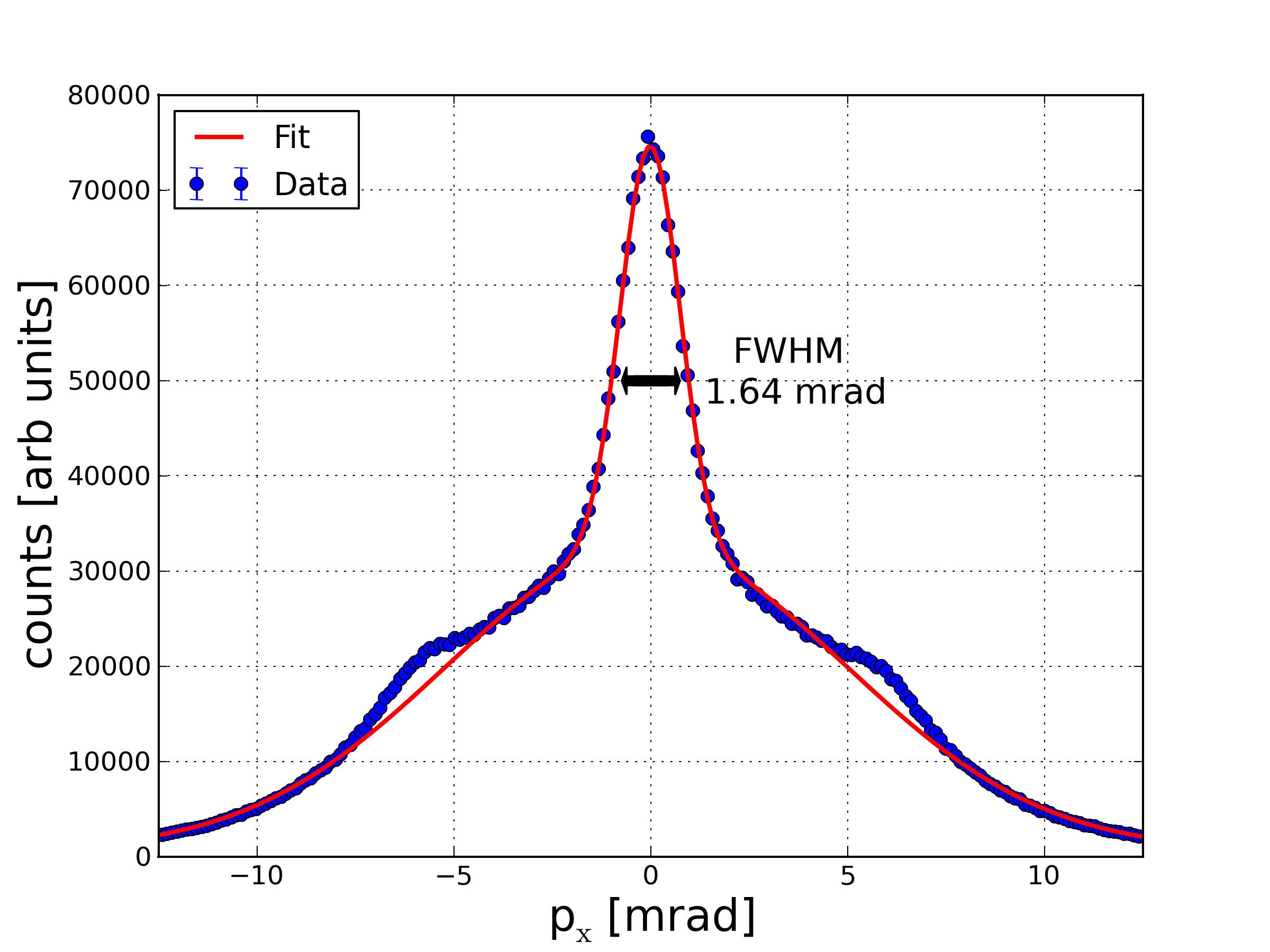} 
  }
    \subfigure[vertical cut]{
    \label{fig:vert}
    \includegraphics[width=0.45\textwidth]{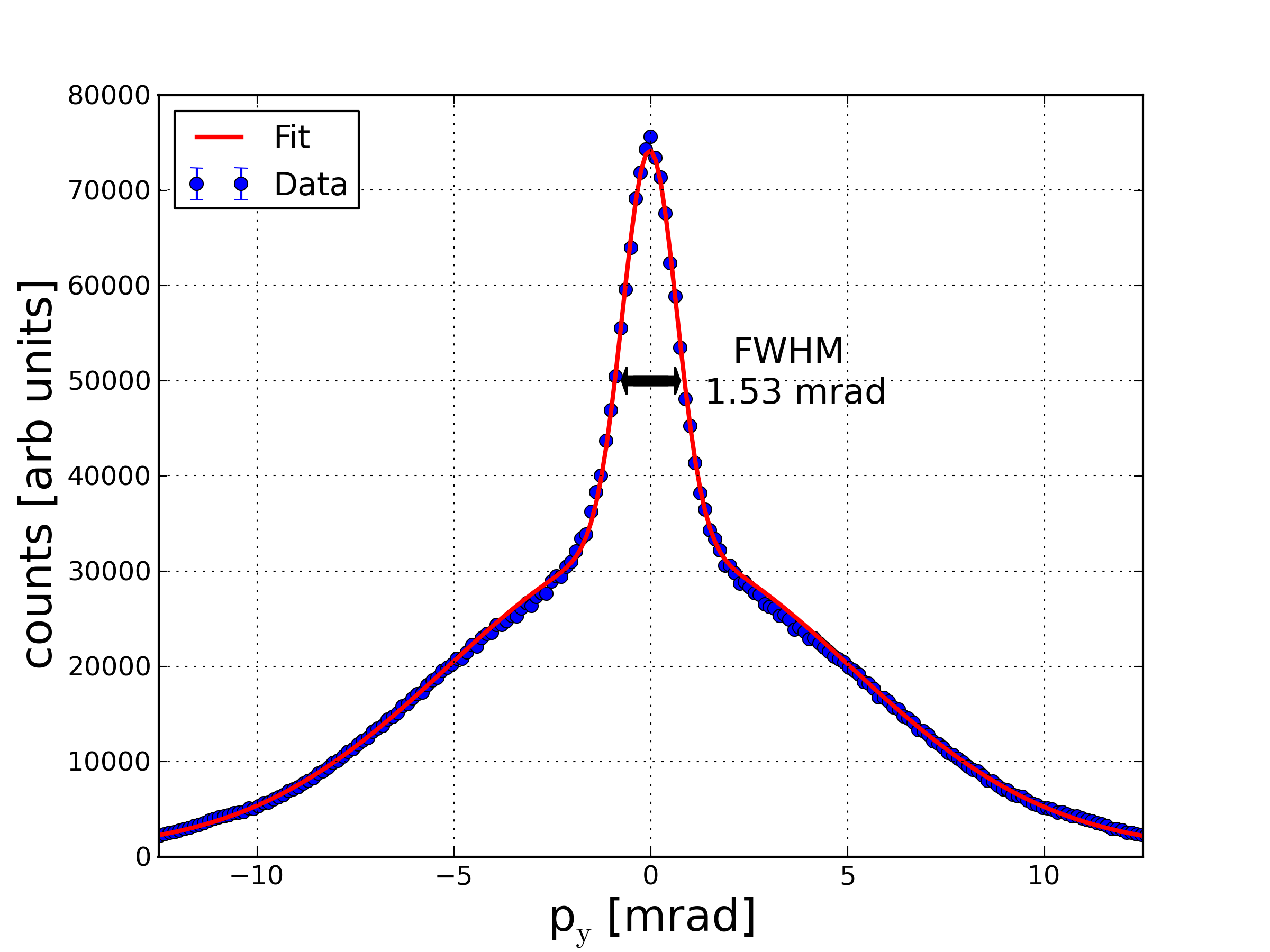} 
  }
  \caption{Cuts through the ACAR spectrum along \ref{fig:hor} $p_{\mathrm{y}}=0$ and \ref{fig:vert} $p_{\mathrm{x}}=0$}
  \label{fig:HorVert}
\end{figure}

A way to directly determine the total angular resolution is the 2D-ACAR measurement of thermalized positronium \cite{Dai91,Berko77}. The angular correlation of its annihilation radiation is only influenced by the thermal momentum of the positronium itself \cite{Ikari79,Kubica75}. For this study  $\alpha$-quartz is used as sample material, because it shows significant formation of positronium, which is thermalized inside the crystal lattice. According to the periodicity of the lattice the positronium wavefunction is therefore represented as a delocalized Bloch-state \citep{Green70}. This can be confirmed in the ACAR measurement since in the momentum domain the higher order momentum components are observable at the positions of the reciprocal lattice vectors (see figure(\ref{fig:quarz}).\\
In the  cross sections of the 2D-ACAR spectrum (figure \ref{fig:HorVert}) two features can be distinguished. The narrow component accords to the para-positronium signal, while the wide component stems from direct positron annihilation and positronium pick-off annihilation. A superposition of two Gaussians is used to describe the data.\\
The width of the narrow component is determined by the angular resolution, which is different for horizontal (x) and vertical (y) direction, since the spot size only contributes to the resolution broadening in the x-directio n(see fig. \ref{fig:hor}). The spot size in x-direction is given by the lateral distribution of the positrons on the sample, while the  spot size in the  y-direction (see fig. \ref{fig:vert}) is determined by the implantation depth of the positrons into the sample, which is of the order of $200\,\mu$m for quarz and therefore can be omitted.\\
Since all contributions to the angular resolution (thermal smearing, spot size, detector resolution) are independent, the total angular resolution is given by the quadratic sum over the individual contributions. 
\[\sigma_{\mathrm{tot}}=\sqrt{ \sigma_{\mathrm{x,y}}^2 + \sigma_{\mathrm{det}}^2 + \sigma_{\mathrm{therm}}^2}\]
Hence, the contribution of the spot size can be extracted, as it is only present the horizontal component. By taking the values given in figure \ref{fig:HorVert} the contribution of the spot size is estimated to be $(0.61\pm 0.13)\,$mrad, which is within the error compatible with the contribution to the resolution that was inferred from the directly measured spot size presented in the previous paragraph. The error of this approximation stems from the statistical accuracy of the fitting procedure. 
%
%
%
%
%
%
%

\section*{Conclusion and Outlook}
A new 2D-ACAR spectrometer has become operational at the Technische Universität München. The above mentioned design goals concerning count rate and resolution were accomplished. We could proof that the contribution to the overall resolution from the spot size is compatible with the measured spot size and therefore well understood. Hence, the next step, the systematic investigation of the electronic structure of correlated materials and the study of temperature driven effects on the electronic structure can be undertaken. Concerning further developments of the spectrometer we focus our efforts on improving the position reconstruction of the detectors. In addition, we plan on upgrading the heatable sample holder, so temperatures up to 800K can be achieved. In the future we plan to move the setup to the new experimental hall of the FRM$\,$II and use the monoenegetic positron beam NEPOMUC to perform depth dependent measurements in order to track the evolution of the electronic structure from the surface to the bulk.

\section*{Acknowledgements}
This project is funded by the Deutsche Forschungsgesellschaft (DFG) within the Transregional Collaborative Research Center TRR 80 ``From electronic correlations to functionality''.

\bibliography{aufbau.bib}

\end{document}